\documentstyle[amsmath,epsf,epsfig,preprint,aps,amssymb]{revtex}
\draft \preprint{SNUTP 04-002}
\begin{document}
\title{\Large\bf Gauges in the bulk}
\author{Jihn E.
Kim$^{(a)}$\footnote{jekim@phyp.snu.ac.kr}, Gary B.
Tupper$^{(b)}$, and Raoul D. Viollier
$^{(b)}$\footnote{viollier@physci.uct.ac.za} }
\address{
$^{(a)}$School of Physics and Center for Theoretical Physics,
Seoul National University, Seoul 151-747,
Korea\\
$^{(b)}$Institute of Theoretical Physics and Astrophysics,
Department of Physics, University of Cape Town, Private Bag,
Rondebosch 7701, South Africa} \maketitle



\def\one{\bf 1}
\def\two{\bf 2}
\def\five{\bf 5}
\def\ten{\bf 10}
\def\tenb{\overline{\bf 10}}
\def\fiveb{\overline{\bf 5}}
\def\threeb{{\bf\overline{3}}}
\def\three{{\bf 3}}
\def\fb{{\overline{F}\,}}
\def\hb{{\overline{h}}}
\def\Hb{{\overline{H}\,}}

\def\To{${\cal T}_1$\ }
\def\Tt{${\cal T}_2$\ }
\def\Z{${\cal Z}$\ }
\def\ot{\otimes}
\def\tr{{\rm tr}}
\def\sinw{{\sin^2 \theta_W}}

\def\slash#1{#1\!\!\!\!\!\!/}

\def\lsl{ l \hspace{-0.45 em}/}
\def\ksl{ k \hspace{-0.45 em}/}
\def\qsl{ q \hspace{-0.45 em}/}
\def\psl{ p \hspace{-0.45 em}/}
\def\ppsl{ p' \hspace{-0.70 em}/}
\def\dsl{ \partial \hspace{-0.45 em}/}
\def\Dsl{ D \hspace{-0.55 em}/}
\def\matrix{ \left(\begin{array} \end{array} \right) }

 \def\NCA{{\em Nuovo Cimento} }
 \def\NIM{{\em Nucl. Instrum. Methods} }
 \def\NIMA{{\em Nucl. Instrum. Methods} A }
 \def\NP{{\em Nucl. Phys.} }
 \def\NPB{{\em Nucl. Phys.} B }
 \def\PL{{\em Phys. Lett.} }
 \def\PLB{{\em Phys. Lett.} B }
 \def\PRL{{\em Phys. Rev. Lett.} }
 \def\PRD{{\em Phys. Rev.} D }
 \def\PR{{\em Phys. Rev.} }
 \def\RMP{{\em Rev. Mod. Phys.} }
 \def\ZPC{{\em Z. Phys.} C }
 \def\PHYSICA{{\em Physica} D }
 \def\CMP{{\em Commun. Math. Phys.} }
 \def\PREP{{\em Phys. Rep.} }
 \def\JMP{{\em J. Math. Phys.} }
 \def\CQG{{\em Class. Quant. Grav.} }
 \def\ANN{{\em Annals of Physics} }
 \def\APP{{\em Acta Phys. Polon.} }
 \def\RPP{{\em Rep. Prog. Phys.} }


\baselineskip 0.8\baselineskip

\vspace{1cm}

\begin{abstract}
We present a general framework for nonparallel brane worlds
and use it to discuss the nonlinear radion problem.
By imposing the Einstein frame as a gauge condition we are
able to give the effective action for both Minkowski and (A)dS$_{4}$
branes. In particular we find the nonlinear radion does not disappear in the second Randall-Sundrum model.

\vskip 0.5cm

\noindent [Key words: brane fluctuation, radion, Randall-Sundrum model]
\end{abstract}

Pacs 11.25.Mj\\
\newpage

\def\one{\bf 1}
\def\two{\bf 2}
\def\five{\bf 5}
\def\ten{\bf 10}
\def\tenb{\overline{\bf 10}}
\def\fiveb{\overline{\bf 5}}
\def\threeb{{\bf\overline{3}}}
\def\three{{\bf 3}}
\def\fb{{\overline{F}\,}}
\def\hb{{\overline{h}}}
\def\Hb{{\overline{H}\,}}

\def\To{${\cal T}_1$\ }
\def\Tt{${\cal T}_2$\ }
\def\Z{${\cal Z}$\ }
\def\ot{\otimes}
\def\tr{{\rm tr}}
\def\sinw{{\sin^2 \theta_W}}

\def\slash#1{#1\!\!\!\!\!\!/}

\def\be{\begin{equation}}
\def\ee{\end{equation}}
\def\ben{\begin{enumerate}}
\def\een{\end{enumerate}}
\def\lsl{ l \hspace{-0.45 em}/}
\def\ksl{ k \hspace{-0.45 em}/}
\def\qsl{ q \hspace{-0.45 em}/}
\def\psl{ p \hspace{-0.45 em}/}
\def\ppsl{ p' \hspace{-0.70 em}/}
\def\dsl{ \partial \hspace{-0.45 em}/}
\def\Dsl{ D \hspace{-0.55 em}/}
\def\matrix{ \left(\begin{array} \end{array} \right) }

\def\ma{m_A}
\def\mf{m_f}
\def\mz{m_Z}
\def\mw{m_W}
\def\ml{m_l}
\def\ms{m_S}
\def\dag{\dagger}

 \def\NCA{{\em Nuovo Cimento} }
 \def\NIM{{\em Nucl. Instrum. Methods} }
 \def\NIMA{{\em Nucl. Instrum. Methods} A }
 \def\NP{{\em Nucl. Phys.} }
 \def\NPB{{\em Nucl. Phys.} B }
 \def\PL{{\em Phys. Lett.} }
 \def\PLB{{\em Phys. Lett.} B }
 \def\PRL{{\em Phys. Rev. Lett.} }
 \def\PRD{{\em Phys. Rev.} D }
 \def\PR{{\em Phys. Rev.} }
 \def\RMP{{\em Rev. Mod. Phys.} }
 \def\ZPC{{\em Z. Phys.} C }
 \def\PHYSICA{{\em Physica} D }
 \def\CMP{{\em Commun. Math. Phys.} }
 \def\PREP{{\em Phys. Rep.} }
 \def\JMP{{\em J. Math. Phys.} }
 \def\CQG{{\em Class. Quant. Grav.} }
 \def\ANN{{\em Annals of Physics} }
 \def\APP{{\em Acta Phys. Polon.} }
 \def\RPP{{\em Rep. Prog. Phys.} }

Tremendous interest has been generated by the Randall-Sundrum 
models
\cite{RS1}
and their generalization from Minkowski to
(Anti) de Sitter branes
\cite{Nih2,Kal3,Kim4}. 
In these models the five dimensional bulk is AdS$_{5}$
resulting in a warp factor which localizes gravity
on the positive tension brane even when the coordinate bulk is made
infinite by putting the negative tension regulator brane on the Cauchy horizon (RSII).
A variant due to Kaloper
\cite{Kal3}
considers only a single dS$_{4}$ brane, the bulk being terminated on the
associated horizon.\\[.2cm]
For two-brane models there exists a scalar mode, the radion, corresponding to
the interbrane distance
\cite{RS1}.
Consequently one obtains
\cite{Gar5}
a scalar-tensor gravity which must respect the constraints provided by 
observation. Within linearized gravity the radion question has been
extensively studied for flat
\cite{Char6,Das7}
as well as bent
\cite{Chac8,Gen9,Gian10}
branes, and such results as the (in)stability of (dS$_{4}$) AdS$_{4}$ branes
established. It has further been found at the linear level that the radion disappears in
the RSII model while in the case of AdS$_{4}$ branes it remains in the
corresponding limit
\cite{Chac8,Gian10}. 
A radion  is also to be expected in Kaloper's model, but to our knowledge this issue
has not been addressed.\\[.2cm]
There are two basic reasons for pursuing the radion problem beyond linear theory.
First, without this one does not know its domain of validity and hence
the reliability of conclusions drawn from it through singular limits. A particular example is the conundrum presented by the disappearance of the RSII radion: putting matter on an isolated brane displaces it from its vacuum position
\cite{Gar5},
and one expects this scalar mode to be reflected in the metric
\cite{Char6}.
Second, it has been suggested that the radion may play an important role in the early universe, as the inflation
\cite{Bar11}
and/or through brane collisions
\cite{Dval12,Khou13,Kansod14},
which necessitates knowing its nonlinear behaviour.\\[.2cm]
Regarding the nonlinear radion results are more scarce: defined as the position relative to a reference brane, Bin\'{e}truy, Deffayet and Langlois 
\cite{Bin15}
have examined the homogeneous radion in general, however the four-dimensional action obtained in this moduli space approach
\cite{Lang16}
is exceedingly awkward due to its nonlinear dependence on the radion velocity.\footnote{Even truncated at quadratic order there is an apparent ghost if the negative tension brane moves, which is resolved by a conformal transformation to the Einstein frame
\cite{Brax17}.}
In the case of Minkowski branes, an effective action has been given by Chiba
\cite{Chib18}
using dimensional reduction which agrees with that of Kanno and Soda
\cite{Kansod19}
who have further derived Kaluza-Klein corrections. Unfortunately, the effective
action of
\cite{Chib18,Kansod19},
which is the basis for the `born again braneworld' model
\cite{Kansod14},
obtains by dimensional reduction from a metric ansatz
\cite{RS1}
that has been criticised 
\cite{Char6}
for failing to solve the linearized equations.
Chiba
\cite{Chib18} has also given an effective action for the improved metric of
\cite{Char6},
but as that metric derives from linear theory the comparison is only
meaningful when implemented in the Einstein frame.
Moreover, these ansatz do not lend themselves to the case of bent branes.\\[.2cm]
In this letter we examine the radion question from a different perspective: restricting to zero
modes, we show that the bulk equations, together with the gauge condition that the effective action be in
the Einstein frame, determines the metric. Our metric, which agrees with
\cite{Char6}
at the linear level, has also been noted by Bagger and Redi
\cite{Bagred20}
in the study of supersymmetric models but without exploring its consequences. Here we give
complete expressions for flat as well as bent branes, including Kaloper's single brane
model. In particular we find the usual disappearance of the radion in RSII is an artifact of linear theory; the general effective action splits into a weak linear coupling regime (which itself disappears in the case of infinite coordinate bulk and for Kaloper's model), an intermediate quadratic coupling and a strong coupling regime.\\[.2cm] 
It is advantageous to choose a coordinate system such that
$g_{(5)\mu 5}=0$ and which admits Einstein spaces of constant
4-curvature,
\begin{eqnarray}
ds^2_{(5)} &=& g_{(5)MN}(X) dX^M dX^N \nonumber\\
&=&\Psi^2(x,y)g_{\mu\nu}(x) dx^\mu dx^\nu -\varphi^2
(x,y)dy^2.  \label{line}
\end{eqnarray}
Then at fixed $x^\mu$, $\int dy \varphi$ gives the distance along the
fifth dimension. By restricting $g_{\mu \nu}$ to be $y$-independent we drop Kaluza-Klein
corrections of order (bulk curvature length)$^{2}$/(four-dimensional wavelengths)$^{2}$
\cite{Kansod19}.
In the Appendix we collected needed tensors.
In particular, $R_{(5)\mu 5}$ is given by
\begin{equation}
R_{(5)\mu 5}=
3\frac{\varphi_{,\mu}}{\varphi}\frac{\Psi^\prime}{\Psi}-3
\left[\frac{\Psi^\prime}{\Psi}\right]_{,\mu}
\end{equation}
The bulk action involves
\begin{eqnarray}
\displaystyle{\int} d^5x \sqrt{g_{(5)}}R_{(5)} & = & \int d^5x \sqrt{-g}\
\Psi^4\varphi R_{(5)} \nonumber\\[.3cm]
                                & = & \Biggl. 
\displaystyle{\int} d^4x \sqrt{-g} \int dy \Biggr\{
R\Psi^2\varphi- 6g^{\alpha\beta}\Psi\varphi\Psi_{,\alpha;\beta}
-2g^{\alpha\beta}\Psi^2\varphi_{,\alpha;\beta} \nonumber\\
                                &   & \Biggl. 
-4g^{\alpha\beta}\Psi\varphi_{,\alpha}\varphi_{,\beta}
+4\frac{\Psi^2}{\varphi} \left[
3(\Psi^\prime)^2+2\Psi\Psi^{\prime\prime}-2\Psi\Psi^\prime
\frac{\varphi^\prime}{\varphi}\right] \Biggr\} \nonumber\\
                                & = & 
\displaystyle{\int} d^4x\sqrt{-g} \displaystyle{\int} dy
\Biggl\{ R\Psi^2 \varphi + 6g^{\alpha \beta}(\Psi \varphi)_{,\alpha}
\Psi_{,\beta} \nonumber\\
                                &   & + 8 \left(
\frac{\Psi^3\Psi^\prime}{\varphi}\right)^\prime
\Biggl. -12\frac{\Psi^2(\Psi^\prime)^2}{\varphi} \Biggr\}
\end{eqnarray}
up to a 4-divergence. The integral of the total $y$ derivative is
cancelled by the Gibbons-Hawking term in the action,
and so may be omitted. Then, with
\begin{equation}
k^2=-\frac{K_{(5)}\Lambda_{(5)}}{6}>0,
\end{equation}
i.e. AdS$_5$,
\begin{eqnarray}
S_{\rm bulk} & = &  \displaystyle{\int} d^5x \sqrt{g_{(5)}} \left[
-\frac{R_{(5)}}{2K_{(5)}}-\Lambda_{(5)}\right] \nonumber\\[.5cm]
&=& \frac{1}{K_{(5)}} \displaystyle{\int} d^4x \sqrt{-g} \displaystyle{\int} dy
\Biggl\{-\frac{R}{2}\Psi^2\varphi- 3g^{\alpha\beta}
(\Psi\varphi)_{,\alpha}\Psi_{,\beta} \nonumber\\[.5cm]
             & &   + 6 \frac{\Psi^2(\Psi^\prime)^2}{\varphi}
            + 6k^2\Psi^4\varphi \Biggr\} 
\end{eqnarray}
It is clear from the Einstein equations outside the
brane(s)
\begin{equation}
G_{(5)MN} \; = \; K_{(5)} \Lambda_{(5)} g_{(5)MN}
\end{equation}
that we must impose a consistency condition
\vspace{-0.2cm}
\begin{equation}\label{cons}
R_{(5) \mu 5} \; = \; 0 \; \; .
\end{equation}
\vspace{-0.2cm}
This leads to
\vspace{-0.2cm}
\begin{equation}
\frac{\Psi^ \prime}{\Psi} = \varphi f
\end{equation}
where $f$ is a function of $y$ only, $f \; = \; f(y) \; .$
So,
\vspace{-0.3cm}
\begin{equation}
\Psi \; = \; {\rm constant} \cdot e^{\int dy \varphi f} \; \; .
\end{equation}
Both $f(y)$ and the constant can be fixed by the
condition that with $\varphi = 1, \; \Psi = W(y)$
obtains for some background warp $W$,
\begin{equation}
\Psi = \exp \left(\int dy \; \varphi \frac{W^ \prime}{W} \right) \; \; .
\end{equation}
Different $\varphi$ specify different
$\lq \lq$gauge choices'' --
essentially different ways of parametrizing distances along the
fifth dimension at fixed $x^{\mu}$ \;.
Note that a brane bound observer cannot actually measure
$\displaystyle{\int} \varphi dy$ which is only seen when viewed from above. Rather
the observer must transmit a signal through the bulk (i.e. some
closed string state) and for such gauge invariant physical observables
all must agree on the result.\\[.2cm]
Equations (8, 10)
are the nonlinear generalization of the condition obtained by
Chacko and Fox
\cite{Chac8}.
In the Randall-Sundrum models $W(y) = e^{-ky}$ on
$0 < y < \ell$, the other region being given by orbifold symmetry; then 
eqs.(8,10) 
are satisfied both by the na\"{i}ve ansatz
\cite{RS1,Chib18,Kansod19}
\begin{equation}
\Psi (x,y) = e^{-ky \varphi (x)} \hspace{.5cm}, 
\hspace{1cm} \varphi (x,y) = \varphi (x)
\end{equation}
and that of Charmousis et al
\cite{Char6,Chib18}
\begin{equation}
\Psi (x,y) = \exp (- ky - \varphi (x) e^{2 k y}) \hspace{.5cm} , \hspace{.5cm} 
\varphi (x,y) = 1 + 2 \; \varphi (x) e^{2 ky} \; \; .
\end{equation}
The gauge 
(11) directly leads via 
eq.(6)
to a Brans-Diche theory in the Jordan frame of the positive tension brane
\cite{Chib18},
$\varphi (x)$ disappearing from the effective action as
$\ell \rightarrow \infty$, the RSII limit.
(Kanno and Soda 
\cite{Kansod19} absorb $\varphi (x)$ and the $l$-dependence
into the Brans-Diche scalar which becomes unity for RSII).
The improved gauge (12)
in 
eq.(6) yields $\varphi (x)$ as a ghost in the RSII model which is resolved by
a conformal transformation to the Einstein frame where it disappears leaving pure
tensor gravity. We credit the latter surprise to the following: linearized theory
is effectively formulated in the Einstein frame whereas the resummation in
eq.(12) places it in the Jordan frame instead 
\cite{Chib18}.
Further, for the 4-dimensional Einstein spaces of constant curvature
\cite{Nih2,Kal3,Kim4}
\begin{equation}
G_{\mu \nu} \; = \; \lambda g_{\mu \nu}
\end{equation}
\begin{equation}
W(y) = \sqrt{\frac{\lambda}{3k^2}} \sinh (kC - k|y|) =
\frac{\sinh (ky_{H} - k|y|)}{\sinh (ky_{H})} \; , \; dS_{4}
\end{equation}
\begin{equation}
W(y) = \sqrt{\frac{- \lambda}{3k^{2}}} \cosh (kC - k|y|) =
\frac{\cosh(kC - k|y|)}{\cosh (kC)} \; , \;  AdS_{4}
\end{equation}
one obtains e.g. for the na\"{i}ve $\varphi (x,y) = \varphi (x)$
\begin{eqnarray*}
\int dy \; \Psi^{2} = \int \; \frac{dx}{k} \frac{e^{- 2 \varphi x}}
{\sqrt{1 + \frac{\lambda}{3k^2} e^{2x}}} 
\end{eqnarray*}
so the integrals cannot be given in closed form.\\[.2cm]
Consider now imposing the gauge condition $\Psi^{2} \varphi = W^{2}$
so the coefficient of $R$ in the effective action is entirely fixed by 
the background and one is automatically in the
Einstein frame. Writing
\begin{equation}
\Psi = \frac{W}{\varphi^{1/2}}
\end{equation}
by differentiating the integral relation between $\varphi$
and $\Psi$
\begin{equation}
\Psi^{\prime} = \frac{W^{\prime}}{\varphi^{1/2}} - 
\frac{W \varphi^{\prime}}{\varphi^{3/2}} = \Psi \varphi
\frac{W^{\prime}}{W} = \frac{W}{\varphi^{1/2}} \varphi
\frac{W^{\prime}}{W}
\end{equation}
or upon rearranging
\begin{equation}
2 \frac{W^{\prime}}{W} = \frac{\varphi^{\prime}}{\varphi (1-\varphi)}
\end{equation}
with solution
\begin{equation}
\frac{W^2}{\phi} = \frac{\varphi}{1 - \varphi} \; , \; \; \phi = \phi(x)
\end{equation}
or
\begin{equation}
\Psi(x,y) = \left[ W^{2} (y) + \phi(x) \right]^{1/2} \; , \; \;
\varphi(x,y) = \frac{W^2(y)}{W^2(y)+\phi(x)} \; .
\end{equation}
This same metric has been given in
\cite{Bagred20}.
At the linearized level 
eqs.(12) and (20) 
agree with
$\phi (x) = - 2 \varphi (x)$ \;.
Here
\begin{eqnarray}
S_{\rm bulk} & = & \frac{1}{K_{(5)}}\int d^4x \sqrt{-g}
\int dy \Biggl\{ -\frac{R}{2}W^2+\frac34 \frac{W^2}{(W^2+\phi)^2}
g^{\alpha\beta}\phi_{,\alpha}\phi_{,\beta} \Biggr. \nonumber\\ 
              &  & \Biggl. + 6\left [k^{2} W^{2} + (W^{\prime})^{2} \right](W^2+\phi) \Biggr\}
\end{eqnarray}
and
\begin{equation}
\int dy\frac{W^2}{(W^2+\phi)^2}=-2\frac{\partial}{\partial
\phi}d_5 \; .
\end{equation}
In this gauge the integrals can be done completely
\cite{Grad21}.
The general result takes the form
\cite{Bagred20}
\begin{eqnarray}
S_{\rm bulk}& = & \int d^4x \sqrt{-g} \; \Biggl[
- \frac{R}{2K} + \frac{3}{4K} \omega (\phi) g^{\alpha \beta}
\phi_{,\alpha} \phi_{,\beta} - \Lambda \Biggr. \nonumber\\
 &   & \Biggl. -(\sigma_0 + \sigma_l) \phi^2 + \sigma_0\Psi^4(x,0)
+ \sigma_l\Psi^4(x,l) \Biggr]
\end{eqnarray}
where the last two terms are cancelled by $S_{\rm brane}$.\\[.2cm]
We now examine the consequences of the gauge
eq.(20)
beginning with the fine-tuned RS models
\cite{RS1}: 
$\Lambda = 0$,
$\sigma_{0} = - \sigma_{l} = 6k / K_{(5)}$,
$W(y) = e^{-k|y|}$.
The relation between $K$ and $K_{(5)}$ is
\begin{equation}
K^{-1} = \frac{2}{K_{(5)}} \; \int_{0}^{l} dy \; W^{2} =
\frac{1 - e^{- 2 k l}}{k \; K_{(5)}}
\end{equation}
and the distance between branes is given by
\begin{equation}
d_{5} = \int_{0}^{l} \; dy \frac{e^{-2ky}}{e^{-2ky} + \phi}
= \frac{1}{2k} \; \ln \left( \frac{1 + \phi}{e^{- 2k l} + \phi} \right) \; \; .
\end{equation}
Note that for all $\phi > 0$, $d_{5}$ remains finite in the RSII 
limit of infinite coordinate bulk, $l \rightarrow \infty$. 
As the induced metrics on the branes are
\begin{equation}
g_{(5) \; \mu \nu} (x, 0) = (1 + \phi (x)) \; g_{\mu \nu} (x) 
\;\; , \;\ g_{(5) \; \mu \nu} (x, l) =
\left( e^{- 2 k l} + \phi (x) \right) \; g_{\mu \nu} (x) \; \; .
\end{equation}
One readily understands this: matter on the positive tension brane 
that would displace it from $y = 0$ in gaussian
normal coordinates here instead is reflected in $\phi \neq 0$ and 
an $x$-dependent distortion of geodesic distances 
\cite{Smol22}. Using eqs.(22,24,25)
\begin{equation}
\omega (\phi) = \frac{1}{(1 + \phi)(e^{- 2kl} + \phi)} \; \; .
\end{equation}
Linear theory takes $\omega (\phi) \simeq \omega (0) = e^{2 k l}$
\cite{Chac8},
whereas eq.(27) shows the limits $\phi \rightarrow 0$ and $l 
\rightarrow \infty$ are not interchangeable. In fact there are three regimes 
which may be evidence by the canonically normalized scalar $\Phi$
\begin{equation}
\frac{3}{4K} \; \omega (\phi) \; g^{\mu \nu} \; \phi_{, \mu} \; \phi_{, \nu} =
\frac{1}{2} \; g^{\mu \nu} \; \Phi_{, \mu } \; \Phi_{, \nu} \; \; ,
\end{equation}
\begin{equation}
\phi = (1 + e^{- 2 k l}) 
\sinh^{2} \; \left( \sqrt{\frac{K}{6}} \Phi \right) +
e^{- k l} \; \sinh \; \left( \sqrt{ \frac{2K}{3} } \; \Phi \right) \; \; .
\end{equation}
Hence, the perturbative regime is $\sqrt{K} \; \Phi \ll 2 \sqrt{6} 
\; e^{- k l} / ( 1 + e^{- 2 k l} )$ where
$\Phi$ couples linearly to matter via the trace of the 4-dimensional 
stress energy tensor but with strength $e^{- k l}$ less than gravitational
\cite{Gar5};
this is absent for infinite coordinate bulk. The intermediate regime
$2 \sqrt{6} \; e^{- k l} \ll \sqrt{K} \Phi \ll 1$ has $\phi \sim K 
\Phi^{2} / 6$ so the
coupling to matter is quadratic and despite being massless 
$\Phi$ produces no long range tree-level two-body forces (one-loop 
2 $\Phi$ exchange generates a fractional charge to the newtonian 
potential of order $K / r^{2}$). The strong coupling region is $\sqrt{K} 
\Phi \gg 1$, i.e. $\phi \gg 1$, which by eq.(25) describes brane 
collisions.\footnote{It has been noted in
\cite{Kansod14}
that the strong coupling in the Einstein frame is conformally equivalent 
to weak coupling in the Jordan frame of observers.}\\[.2cm]
Next we consider the detuned models 
\cite{Nih2,Kal3,Kim4}
where $\ell$ is not a free parameter,
taking the AdS$_{4}$ case of eq.(15) as an example.
Here $\Lambda = \lambda / K < 0$, $\sigma_{0} = (6k / K_{(5)})\; \tanh (k C)$,
$\sigma_{l} = (6k / K_{(5)}) \tanh (k l - k C) > - \sigma_{0}$ 
so the potential term in eq.(23) is stable
\cite{Chac8,Bagred20}.
The $K - K_{(5)}$ relation and $d_{5}$ are given by
\begin{equation}
K^{-1} = \frac{1}{k K_{(5)}} \; \left[ \coth (kC) + sec h^{2} (kC) \;
\left\{ k l + \sinh (k l - k C) \cosh ( k l - k C) \right\} \right] \; \; ,
\end{equation}
\begin{eqnarray}
d_5 & = & l - \displaystyle{ \frac{1}{k \sqrt{1 + 
\displaystyle { \frac{1}{\phi \cosh^2(kC)}}}  } 
\left\{ {\rm arccoth \left(
\sqrt{1 + \frac{1}{\phi \cosh^2(kC)}} \coth(kC)
\right) } \right\} } \nonumber\\[1cm] 
    &   & + \; {\rm arccoth} \left(
\sqrt{ 1 +  \displaystyle{ \frac{1}{\phi \cosh^2(kC)}} } \; \coth\;(kl - kC)
\right) \; \; ,
\end{eqnarray}
respectively, and then $\omega (\phi)$ obtains via eq.(22). 
It is more useful, however, to note that observation restrict a 
bonafide cosmological constant to
$|\lambda|^{1/2} < H_{0}$ where $H_{0} \sim 10^{-42}$ GeV is the present 
value of the Hubble constant, while $k >$ meV from submilimeter gravity 
experiments; thus ${\rm sech}(kC) < H_{0}/k \sim 10^{- 30}$. Taking the limit 
$kC \rightarrow \infty$ in eqs.(30,31) reproduces eqs.(24,25), and
$\sigma_{0} + \sigma_{l} \simeq (6 k/K_{(5)}) \; \sec h^{2} (kG)/[1 - 
\coth (k l) ]$ so the effective action is
\begin{eqnarray*}
S_{\rm eff} \; \simeq \; \int \; d^{4} x \; \sqrt{-g} 
\; \Biggl[ - \frac{R}{2K} + \frac{3}{4K} \;
\frac{g^{\mu \nu} \; \phi_{, \mu} \; \phi_{, \nu}}{(1 + \phi) 
(e^{- 2 k l} + 4)} \; - \; \Lambda \Biggr. 
                   \Biggl. \; + \; \Lambda \; e^{2 k l} \; 
\phi^{2} \; + {\cal{L}_{\it m}} \Biggr]  
\end{eqnarray*}
where the matter lagrangian $\cal{L}_{\it m}$ is constructed using eq.(26). 
By eq.(29) one sees that $\Phi$ acquires a mass $m_{\Phi} = \sqrt{- 4 \; K 
\Lambda / 3}$ while $m_{\Phi}^{-1}$ is horizon size.\\[.2cm]
Albeit eq.(32) was obtained for AdS$_{4}$, similar steps lead to the same 
expression in the case of de Sitter branes -- there $\Lambda > 0$ so the 
potential is unstable. More generally, for two branes with slow-roll fields, 
eq.(32) holds so long as $H \ll k$ which for the expectation $k \sim K_{
(5)}^{- 1/2} \sim K^{- 1/2} \sim 10^{18}$ GeV is rather loose; for larger 
$H$ one should revert to a five-dimensional description to reflect that $H 
\sim \rho$ rather than $H \sim \sqrt{\rho}$ as in the four-dimensional 
effective action \cite{Lang16}.\\[.2cm]
Finally we consider Kaloper's model, consisting of a single positive 
tension brane \cite{Kal3}. Here
\begin{equation}
K^{-1} \; = \; \frac{2}{K_{(5)}} \; \int_{0}^{y_{H}} \; dy \; W^{2} (y) =
\frac{1}{k K_{(5)}} \; \left[ \coth\;(k y_{H}) - ky_{H} \; 
{\rm cosech}^{2} \; (k y_{H}) \right]
\end{equation}
from eq.(14), and
\begin{equation}
d_{5} = y_{H} - \frac{1}{k \sqrt{1 - 
\displaystyle{   \frac{1}{\phi \sin h^{2} (ky_H)}}}   }
{\rm arctanh}\; \left(
\sqrt{1 - \frac{1}{\phi \sin h^2(ky_H)}} 
\tanh\;(ky_{H}) \right)
\end{equation}
is the distance to the Cauchy horizon $W(y_{H}) = 0$. For\footnote{We omit 
there the region $\phi \ll \; {\rm cosech}^{2} \; (k y_{H})$ where
$d_{5} \simeq y_{H}$ - $\frac{\pi}{2k}$ $\sinh \; (ky_{H})$  $\sqrt{\phi}$ 
and $\omega (\phi) \simeq \frac{\pi}{2}$ 
$\frac{\sinh \; (k y_{H})}{\sqrt{\phi}}$ as it is tiny for cosech 
$(k y_{H} ) < 10^{- 30}$.} $H \ll k$, 
$k y_{H} \gg 1$
\begin{equation}
K \simeq k K_{(5)}
\end{equation}
\begin{equation}
d_{5} \simeq \frac{1}{2k} \; \ln \left( \frac{1 + \phi}{\phi} \right)
\end{equation}
and $\sigma_{0} = (6 k/K_{(5)}) \coth (k y_{H}) \simeq 6 k^{2}/K$ so
\begin{equation}
S_{\rm eff} \simeq \int \; d^{4} x \; \sqrt{- g} \; \left[ - \frac{R}{2K} 
+ \frac{3}{4K} \;
\frac{g^{\mu \nu} \; \phi_{, \mu} \; \phi_{, \nu}}{\phi (1 
+ \phi)} - \Lambda - 6 k^{2} \phi^{2} \; \frac{6 k^{2}}{K} \; 
+ \cal{L}_{\it m} \right]
\end{equation}
The distinctive features of this model are:
\begin{itemize}
\item[(i)] the absence of a perturbative radion regime, $\omega ( \phi 
\rightarrow 0) \rightarrow \infty$, despite which the canonical scalar $\Phi$
\begin{equation}
\phi = \sinh^{2} \; \left( \sqrt{\frac{K}{6}} \; \Phi \right)
\end{equation}
is well behaved, being massless, quadratically 
coupled to matter and quartically self-coupled for $\sqrt{K} \; \Phi \ll 1$;
\item[(ii)]
the potential is stable whereas two dS$_{4}$ branes are unstable -- 
this occurs because the hidden brane with tension $\sigma_{l} < - 
\sigma_{0}$ is absent, being replaced by the horizon.\\[.2cm]
\end{itemize}
In conclusion, we have re-examined the nonlinear radion problem at the 
zero-mode level while imposing the gauge condition that the dimensionally 
reduced effective action be in the Einstein frame. Our main results 
for two-brane models, eqs.(32, 29) establish that the radion in RSII 
does not disappear but rather changes its character. This is in broad 
agreement with expectations \cite{Char6,Smol22} 
and in contrast to previous formulations \cite{Chib18}.
Generally, the scalar $\phi$ in our effective action
eq.(32) exhibits three regimes depending on its value compared to 
$e^{- 2 k l}$ or 1. We have also obtained the effective action for Kaloper's 
one-brane model which provides an exception to the rule that dS$_{4}$ 
branes are unstable. 

\vspace{1.5cm}

\centerline{\bf Acknowledgments}
One of us (JEK) is supported in part by the KOSEF ABRL Grant No. 
R14-2003-012-01001-0, and the BK21 program of
Ministry of Education. Two of us (RDV, GBT) acknowledge grants
from the South African National Research Foundation (NRF
GUN-2053794), the Research Committee of the University of Cape
Town and the Foundation for Fundamental Research (FFR
PHY-99-01241).

\newpage

\centerline{\bf Appendix}

In this appendix, we give expressions of Ricci tensor
and Ricci scalar with the metric of eq.(1).
The Christoffel symbol is
\begin{equation}
\Gamma_{(5)MN}^A \; = \; \frac12
 \; g_{(5)}^{AB} \; \left[ g_{(5)MB,N}+g_{(5)NB,M}-g_{(5)MN,B} \right] \; .
\end{equation}
Explicitly, they are 
\begin{eqnarray}
\Gamma_{(5)\mu \nu}^{\alpha} & = & \Gamma_{\mu\nu}^{\alpha}
+ g^{\alpha}_{\mu} \frac{\Psi_{,\nu}}{\Psi}
+ g^{\alpha}_{\nu} \frac{\Psi_{,\mu}}{\Psi}
- g^{\alpha \beta} \; g_{\mu \nu} \frac{\Psi_{, \beta}}{\Psi}\\[.2cm]
\Gamma_{(5) \mu\nu}^{5}      & = & g_{\mu\nu} \frac{\Psi \Psi^{\prime}}{\varphi^2}\\[.2cm]
\Gamma_{(5) \mu 5}^{\alpha}  & = & g^{\alpha}_{\mu} \frac{\Psi^{\prime}}{\Psi}\\[.2cm]
\Gamma_{(5) 55}^{\alpha}     & = & g^{\alpha \beta} \frac{\varphi \varphi_{, \beta}}{\Psi^2}\\[.2cm]
\Gamma_{(5) \mu 5}^{5}       & = & \frac{\varphi_{, \mu}}{\varphi}\\[.2cm]
\Gamma_{(5)55}^{5}           & = & \frac{\varphi^{\prime}}{\varphi} \; .
\end{eqnarray}
Contractions give
\begin{eqnarray}
\Gamma_{(5) \mu A}^{A} & = &\Gamma_{\mu\alpha}^\alpha+4\frac{\Psi_{,\mu}}{\Psi}
+ \frac{\varphi_{,\mu}}{\varphi}\\[.3cm]
\Gamma_{(5) 5 A}^{A}   & = & 4 \frac{\Psi^{\prime}}{\Psi}
+ \frac{\varphi^{\prime}}{\varphi} \; .
\end{eqnarray}
Then from
\begin{eqnarray}
R_{(5)MN} = \Gamma_{(5) MN, A}^{A} - \Gamma_{(5) MA, N}^{A}
+ \Gamma_{(5) MN}^{A} \Gamma_{(5) AB}^{B}
- \Gamma_{(5) MA}^{B} \Gamma_{(5) NB}^{A}
\end{eqnarray}
we obtain the Ricci tensors,
\begin{eqnarray}
R_{(5) \mu 5}& = &
\Gamma_{(5) \mu 5, \alpha}^{\alpha} + \Gamma_{(5) \mu 5, 5}^{5}
- \Gamma_{(5) \mu A, 5}^{A}
+ \Gamma_{(5) \mu 5}^{\alpha} \Gamma_{(5) \alpha B}^{B}
+\Gamma_{(5) \mu 5}^{5} \Gamma_{(5) 5 B}^{B} \nonumber\\[.3cm]
             &   & - \Gamma_{(5) \mu \alpha}^{\beta} \Gamma_{(5) 5\beta}^{\alpha}
-\Gamma_{(5) \mu \alpha}^{5} \Gamma_{(5) 55}^{\alpha}
-\Gamma_{(5) \mu 5}^{\beta} \Gamma_{(5) 5\beta}^{5}
-\Gamma_{(5) \mu 5}^{5} \Gamma_{(5) 55}^{5} \nonumber\\[.3cm]
             & = & \left[ \frac{\Psi^{\prime}}{\Psi} \right]_{, \mu} 
+ \left[ \frac{\varphi_{, \mu}}{\varphi} \right]_{, 5} 
- \left[ 4 \frac{\Psi_{,\mu}}{\Psi}
+ \frac{\varphi_{,\mu}}{\varphi} \right]_{,5}
+ \frac{\Psi^{\prime}}{\Psi} \Gamma_{(5) \mu B}^{B} \nonumber\\[.3cm]
            &   & + \frac{\varphi_{, \mu}}{\varphi} \left[
4 \frac{\Psi^{\prime}}{\Psi} + \frac{\varphi^{\prime}}{\varphi} \right]
- \frac{\Psi^{\prime}}{\Psi} \Gamma_{(5) \mu \alpha}^{\alpha}
- \left[ g_{\mu \alpha} \frac{\Psi \Psi^{\prime}}{\varphi^2} \right]
  \left[ g^{\alpha \beta} \frac{\varphi \varphi_{,\beta}}{\Psi^{2}} \right] \nonumber\\[.3cm]
            &    & - \frac{\Psi^{\prime}}{\Psi} \Gamma_{(5) \mu 5}^{5} 
- \left[ \frac{\varphi_{, \mu}}{\varphi} \right]
  \left[ \frac{\varphi^{\prime}}{\varphi} \right] \nonumber\\[.3cm]
            & = & 3 \frac{\varphi_{, \mu}}{\varphi} \frac{\Psi^{\prime}}{\Psi}
- 3 \left[ \frac{\Psi^{\prime}}{\Psi} \right]_{, \mu} \; .
\end{eqnarray}

\vspace{.5cm}

Similarly, we obtain
\begin{eqnarray}
R_{(5)5 5}&=&
\Gamma_{(5) 5 5,\alpha}^\alpha+ \Gamma_{(5) 5 5,5}^5
- \Gamma_{(5) 5 A,5}^A
+\Gamma_{(5) 5 5}^\alpha \Gamma_{(5) \alpha B}^B
+\Gamma_{(5) 5 5}^5 \Gamma_{(5) 5 B}^B \nonumber \\[.2cm]
&&-\Gamma_{(5) 5\alpha}^\beta \Gamma_{(5) 5\beta}^\alpha
-\Gamma_{(5) 5\alpha}^5 \Gamma_{(5) 55}^\alpha
-\Gamma_{(5) 5 5}^\beta \Gamma_{(5) 5\beta}^5
-\Gamma_{(5) 5 5}^5 \Gamma_{(5) 55}^5 \nonumber \\[.2cm]
&=&
\left[{g^{\alpha\beta}}\frac{\varphi\varphi_{,\beta}}{\Psi^2}
\right]_{,\alpha}+\left[\frac{\varphi^\prime}{\varphi}\right]_{,5} 
-\left[4\frac{\Psi^\prime}{\Psi}+\frac{\varphi^\prime}{\varphi}
\right]_{,5} \nonumber \\[.2cm]
&&
+\left[g^{\alpha\beta}\frac{\varphi\varphi_{,\beta}}{\Psi^2}\right]
\left[\Gamma_{\alpha\beta}^\beta+4\frac{\Psi_{,\alpha}}{\Psi}
+\frac{\varphi_{,\alpha}}{\varphi}\right]
+\left[\frac{\varphi^\prime}{\varphi}\right]\left[
4\frac{\Psi^\prime}{\Psi}+\frac{\varphi^\prime}{\varphi}\right] \nonumber \\[.2cm]
&&-4\left[\frac{\Psi^\prime}{\Psi}\right]^2
-2\left[\frac{\varphi_{,\alpha}}{\varphi}\right]
\left[g^{\alpha\beta} \frac{\varphi\varphi_{,\beta}}{\Psi^2}\right]
-\left[\frac{\varphi^\prime}{\varphi}\right]^2 \nonumber \\[.2cm]
&=&
g^{\alpha\beta}\frac{\varphi\varphi_{,\beta;\alpha}}{\Psi^2}
+2g^{\alpha\beta}\varphi
\frac{\Psi_{,\alpha}\varphi_{,\beta}}{\Psi^3}
-4\frac{\Psi^{\prime\prime}}{\Psi}
+4\frac{\varphi^\prime}{\varphi}\frac{\Psi^\prime}{\Psi}
\end{eqnarray}

\begin{eqnarray}
R_{(5)\mu \nu}&=&
\Gamma_{(5) \mu \nu,\alpha}^\alpha+ \Gamma_{(5) \mu \nu,5}^5
- \Gamma_{(5) \mu A,\nu}^A
+\Gamma_{(5) \mu \nu}^\alpha \Gamma_{(5) \alpha B}^B
+\Gamma_{(5) \mu \nu}^5 \Gamma_{(5) 5 B}^B \nonumber \\[.2cm]
&&-\Gamma_{(5) \mu\alpha}^\beta \Gamma_{(5) \nu\beta}^\alpha
-\Gamma_{(5) \mu\alpha}^5 \Gamma_{(5) \nu 5}^\alpha
-\Gamma_{(5) \mu 5}^\beta \Gamma_{(5) \nu \beta}^5
-\Gamma_{(5) \mu 5}^5 \Gamma_{(5) \nu 5}^5 \nonumber \\[.2cm]
&=&
R_{\mu\nu}-2\frac{\Psi_{,\mu;\nu}}{\Psi} -g_{\mu\nu}
g^{\alpha\beta}\left[\frac{\Psi_{,\alpha;\beta}}{\Psi}
+\frac{\Psi_{,\alpha}\Psi_{,\beta}}{\Psi^2}\right] \nonumber \\[.2cm]
&&-\frac{\varphi_{,\mu;\nu}}{\varphi}
+4\frac{\Psi_{,\mu}\Psi_{,\nu}}{\Psi^2}
+\frac{\varphi_{,\mu}\Psi_{,\nu}+\Psi_{,\mu}\varphi_{,\nu}}{\varphi\Psi} \nonumber \\[.2cm] 
&& -g_{\mu\nu}g^{\alpha\beta}\frac{\varphi_{,\alpha}\Psi_{,
\beta}}{\varphi\Psi}+\frac{g_{\mu\nu}}{\varphi^2}
\left[3(\Psi^\prime)^2+\Psi\Psi^{\prime\prime}-\Psi\Psi^\prime
\frac{\varphi^\prime}{\varphi}\right] \; .
\end{eqnarray}

\vspace{.5cm}

The Ricci scalar is
\begin{eqnarray}
R_{(5)} &=& \Psi^{-2}\Big\{ R-6g^{\alpha\beta}\frac{\Psi_{,\alpha;\beta}}{
\Psi}-4g^{\alpha\beta}\frac{\Psi_{,\alpha}\Psi_{,\beta}}{\Psi^2}
-g^{\alpha\beta}\frac{\varphi_{,\alpha;\beta}}{\varphi} \nonumber \\[.2cm] 
&&+4g^{\alpha\beta}\frac{\Psi_{,\alpha}\Psi_{,\beta}}{\Psi^2}
-2g^{\alpha\beta}\frac{\varphi_{,\alpha}\Psi_{,\beta}}{\varphi\Psi}
+\frac{4}{\varphi^2}\left[3(\Psi^\prime)^2+\Psi\Psi^{\prime\prime}
-\Psi\Psi^\prime\frac{\varphi^\prime}{\varphi}\right]\Big\} \nonumber \\[.2cm] 
 &&-\frac{1}{\varphi^2}\Big\{g^{\alpha\beta}\frac{\varphi}{\Psi^2}
\varphi_{,\beta;\alpha} +2g^{\alpha\beta}\varphi\frac{\Psi_{,\alpha}
\varphi_{,\beta}}{\Psi^3}
-4\frac{\Psi^{\prime\prime}}{\Psi}+4\frac{\varphi^\prime}{\varphi}
\frac{\Psi^\prime}{\Psi} \Big\} \; .
\end{eqnarray}

\end{document}